\documentclass[twocolumn]{aastex631}

\usepackage{xcolor}
\usepackage{amsmath,amsthm,amsfonts}
\usepackage{graphicx}
\usepackage{physics}
\usepackage{hyperref}
\hypersetup{
	colorlinks   = true, %Colours links instead of ugly boxes
	urlcolor     = blue, %Colour for external hyperlinks
	linkcolor    = blue, %Colour of internal links
	citecolor    = blue   %Colour of citations
}

\usepackage{diagbox}
\graphicspath{{figs/}}

\newcommand{\N}{{\mathcal{N}}}

\newcommand{\nc}{\newcommand}

\nc{\calR}{{\cal{R}}}
\nc{\calP}{{\cal{P}}}
\nc{\cN}{ {\cal{N}} }

\def\bfx{{\bf x}}

\newcommand{\Mp}{M_{_{\rm Pl}}}
\nc{\Mpt}{M_{_{\rm Pl}}^2}

\begin{document}

\title{Inflation from Multiple Pseudo-Scalar Fields:\\
	PBH Dark Matter and Gravitational Waves}

\author[0000-0002-2807-404X]{Alireza Talebian}
\affiliation{School of Astronomy, Institute for Research in Fundamental Sciences (IPM), Tehran, Iran, P.O. Box 19395-5531}
%\affiliation{Center for Gravitational Physics and Quantum Information,\\
%	Yukawa Institute for Theoretical Physics (YITP), Kyoto University, Kyoto 606-8502, Japan}

\author[0000-0003-0641-6237]{Seyed Ali Hosseini Mansoori}
\affiliation{Faculty of Physics, Shahrood University of Technology, P.O. Box 3619995161 Shahrood, Iran}

\author[0000-0002-1850-4392]{Hassan Firouzjahi}
\affiliation{School of Astronomy, Institute for Research in Fundamental Sciences (IPM), Tehran, Iran, P.O. Box 19395-5531}

\begin{abstract}
We study a model of inflation with multiple pseudo-scalar fields coupled to a $U(1)$ gauge field through Chern-Simons interactions. Because of parity violating interactions, one polarization of the gauge field is amplified yielding to enhanced curvature perturbation power spectrum. Inflation proceeds in  multiple stages as each pseudo-scalar field rolls towards its minimum yielding to distinct multiple peaks in the curvature perturbations power spectrum at various scales during inflation. The localized peaks in power spectrum generate  Primordial Black Holes (PBHs) which can furnish a large fraction of Dark Matter (DM) abundance.  In addition,  gravitational waves (GWs) with non-trivial spectra are generated which are in sensitivity range of various forthcoming GW observatories. 
\end{abstract}


\section{Introduction}
\label{sec:intro}
Inflation is the leading paradigm for early universe cosmology and the mechanism behind the generation of large scale structures. Among the basic predictions of models of inflations are that the primordial perturbations are nearly scale invariant, adiabatic and Gaussian which are  
well consistent with cosmological observations \citep{Planck:2018jri}. While the simplest models of inflation are based on a single scalar field, having inflation driven by multiple scalar fields along with other types of fundamental fields in the spectrum  is well-motivated in  models of high energy physics \citep{Weinberg:2008zzc, Wands:2007bd,  Baumann:2009ds}. In particular, 
there have  been growing interests in the axion models~\citep{McAllister:2008hb, Anber:2009ua,Barnaby:2011vw,Barnaby:2010vf,Durrer:2010mq, Bugaev:2013fya,Linde:2012bt,Garcia-Bellido:2016dkw,Talebian:2022jkb} to amplify the primordial power spectrum for PBHs formation and to generate detectable GWs signals.

PBHs are distinct from their astrophysical counterparts in several ways. Among all, PBHs could form in the early universe from the collapse upon horizon re-entry of perturbations generated during inflation which may comprise a large fraction of DM energy density 
\citep{carr1975primordial,Carr:2009jm,Carr:2020xqk,Carr:2016drx,Sasaki:2018dmp}. However, PBHs can form through different channels in the early universe as well~\citep{Khlopov:2008qy,Carr:2020gox}. Remarkably, unlike astrophysical black holes, PBHs can include a vast range of masses. Therefore, the recent observations of GWs from merging binary systems with about 30 times solar mass
~\citep{LIGOScientific:2016aoc} together with the lack of observational signals of particle DM have renewed the interests in PBHs from inflation \citep{Bird:2016dcv,Clesse:2016vqa,Sasaki:2016jop}.
To produce PBHs  from inflation one requires that the amplitude of the primordial curvature perturbation is large enough, at least $10^{7}$ times larger than 
its  CMB value. Over the past a variety of single field models have been studied  to provide such enhancement, for a review see \citep{Sasaki:2016jop, Green:2020jor, Byrnes:2021jka} and the references therein.

Among multiple-field inflation scenarios $\mathsf N$-flation is an interesting 
example which is based on many axion fields,  %coupled to a $U(1)$ gauge fields, 
providing  a simple radiatively stable realization of chaotic inflation \citep{Dimopoulos:2005ac}. In this model, the collective contributions of $\mathsf N$ axion fields yield  a long enough period of inflation to solve the flatness and horizon problems. In this picture, inflation is divided to $\mathsf N$ slow-roll phases where each phase is driven by one axion while others are nearly frozen. Inspired by $\mathsf N$-flation model, in this work,  we study an inflationary model with multiple pseudo-scalar fields coupled to a $U(1)$ gauge field through Chern-Simons types of interaction. We examine the enhancement of the curvature power spectrum to form PBHs at small scales. We show that PBHs can be formed abundantly (in the allowed window where PBHs could provide a substantial part of the DM, if not all) without introducing specific features on the inflationary potentials. In addition, tensor perturbations with non-trivial spectrum are generated  which may be detected in  upcoming GWs experiments.

\section{The Model and Background Dynamics} \label{sec:model}
We consider $\sf N$ pseudo-scalar fields $\Phi_a$ ($a=1,2, \cdots, \sf N$) driving inflation in $\sf N$ stages. While our starting discussions are general but for specific examples studied below, we consider the cases $\sf N =2, 3$ specifically. 
In each stage, only one pseudo-scalar field can slow-roll and then decay, while others remain frozen. The next inflationary stage is driven by the second field before it  decays and so on. For this picture to be realized, we need a working hierarchy on the masses of $\Phi_a$, such that the most massive field starts rolling first, then the second most massive field and so on \citep{Yokoyama:2007dw}. 
{For example if the ratio of the mass of $\Phi_1$ to $\Phi_2$ is at the order $10$ or so, then we can safely assume that the first period of inflation is driven by $\Phi_1$.} 
As in single field axion model  we demand that all pseudo-scalar fields couple to a $U(1)$ gauge field $A_\mu$ through the Chern-Simons interactions~\citep{anber2006n,Bachlechner:2019vcb} in the following action
\begin{eqnarray}
	\label{action}
	\nonumber S &=& \int \dd^4 x \, \sqrt{-g}\bigg[ \dfrac{\Mp^2}{2} R 
	- \frac{1}{2} \delta^{ab}  g^{\mu \nu} \partial_{\mu} \Phi_{a}\partial_{\nu} \Phi_{b}-V(\Phi_{a})\\
	&-&\frac{1}{4}F_{\mu\nu}F^{\mu\nu}-\dfrac{1}{4}\sum_{a=1}^{\sf N}\tilde{\alpha}_a \Big(
	\dfrac{\Phi^{a}}{\Mp}
	\Big)F_{\mu\nu}\tilde{F}^{\mu\nu}\bigg] \,,
\end{eqnarray}
in which $\Mp$ is the reduced Planck mass, $R$ is the Ricci scalar associated with the spacetime metric $g_{\mu\nu}$. In addition, $\tilde{F}^{\mu\nu}\equiv \epsilon^{\mu\nu\rho\sigma} F_{\rho\sigma}/2$ is the dual of the gauge field strength tensor  $F_{\mu\nu} = \nabla_\mu A_\nu - \nabla_\nu A_\mu$. %\textcolor{orange}{(Notation of \citep{Barnaby:2011vw}, below Eq. (2.1))} 
Finally, $\tilde{\alpha}_{a}$ is a dimensionless parameter controlling the  coupling of the $a$-the pseudo-scalar field to the electromagnetic field~\citep{anber2006n}.
To simplify the analysis below, we assume that all $\tilde{\alpha}_{a}$ have the same sign. This is a technical tuning which simplifies the analysis significantly but it can be relaxed in a more general consideration.

In the presence of  coupling $\tilde{\alpha}_{a}$, the gauge field quanta exhibit tachyonic instability sourced by the rolling pseudo-scalar fields. More precisely, during the $a$-th stage of inflation, only the  field  $\Phi_a$ rolls slowly. The rolling  of 
this $\Phi_a$ amplifies one polarization (e.g. the negative-helicity) of the gauge field, leading to~\citep{Anber:2006xt}
\begin{align}
	\label{amplified_A}
	A_{k}^{(-)}\!\cong\!\dfrac{e^{\pi \xi-\sqrt{8\xi k/({\bf a} H)}}}{\sqrt{2k}}\!\Big( \dfrac{k}{2\xi {\bf a} H} \Big)^{\frac{1}{4}}\,;
	\hspace{.38cm}
	\xi\equiv  \!\sum_{a=1}^{\sf N} \dfrac{|\tilde{\alpha}_a \dot{\phi}^a|}{2H} \, ,
\end{align}
%{\color{red}{Is $\xi$ the sum over multiple stage or given at each stage separately?}}
where $k$ is the comoving Fourier mode of the gauge field, ${\bf a}$ and $H\equiv \dot {\bf a}/{\bf a}$ respectively are the scale factor and the Hubble expansion rate during inflation and $\phi_{a}$ is the homogeneous part of the pseudo-scalar field $\Phi_a$. The dot denotes the derivative with respect to the cosmic time. The above solution well describes the growth of the mode functions in the interval $(8\xi)^{-1} \lesssim k/(aH) \lesssim 2\xi $~\citep{Barnaby:2010vf}. Note also that the other polarization state (here the positive-helicity) is not amplified and can therefore be ignored. The so-called \textit{instability parameter} $\xi$ can be considered nearly constant, as its time variation is subleading in a slow-roll expansion. 
It is worth mentioning that the gauge quanta \eqref{amplified_A} not only affect the background dynamics of  $\phi_{a}$ and the scale factor but also source scalar perturbations via \textit{inverse decay}~\citep{Barnaby:2011vw,Barnaby:2010vf, Barnaby:2011qe}.
We assume  that $\dot{\phi}_a<0$ during inflation as in the large field models like \eqref{eq:potential} so for $\tilde{\alpha}_a>0$  the negative-helicity is amplified. 
{As mentioned before, we assume that all  $\tilde{\alpha}_{a}$ are positive so only $A_{k}^{(-)} $ is amplified at each stage of inflation.}

Since the gauge field has no background value one can calculate their effects on the background dynamics via mean field approximation method~\citep{Linde:2012bt,Talebian:2022jkb}, yielding to
\begin{eqnarray}
	\label{Friedmann}
	3 \Mp^2 H^2 - V - \dfrac{1}{2}\delta^{ab}\dot{\phi}_a\dot{\phi}_b%(\phi^a)
	= \boldsymbol{\rho_{em}} &\simeq&\frac{\Gamma(7)}{2^{19}\pi^2}\frac{H^4}{\xi^3}  e^{2\pi \xi} \,,
	\\
	\label{KG}
	\ddot{\phi}_a+ 3\,H\,\dot{\phi}_a+\frac{\partial V}{\partial \phi_a}= \boldsymbol{J_{a}}&\simeq& -\frac{
		\tilde{\alpha}_a\Gamma(8)}{2^{21}\pi^2} \frac{H^4}{\xi^4} e^{2\pi\xi} 	
	\,.
\end{eqnarray}
Note that the contributions in right hand sides above come respectively from $\rho_{em} \propto \langle E^2+B^2 \rangle$ and $J_{a} \propto \tilde{\alpha}_{a}\langle E\cdot B \rangle$ where the electric and magnetic fields, in the Coulomb-radiation gauge $(A_0 = 0 = \partial_i A^i)$, are defined as $E^i \equiv -{\bf a}^{-1}\dot{A}^i$ and $B^i \equiv {\bf a}^{-2} \epsilon^{ijk}\partial_j A_k$ respectively. The exponential enhancement reflects significant non-perturbative gauge particle production in the regime $\xi \gtrsim 1$~\citep{Anber:2009ua}. 
To ensure that the tachyonic growth of gauge field fluctuations does not spoil the inflationary dynamics, we  demand
$H^2/|\dot \phi^a| \ll {\cal O}(10^2) ~\xi^{3/2}e^{-\pi \xi}$ at each $a$-th stage~\citep{Barnaby:2010vf,Barnaby:2011vw,Talebian:2022jkb}. 
From Eq.~\eqref{KG}, one finds that the growth  of $\xi$ comes to a halt when the back-reaction term becomes large enough. Note that the system experiences a nonlinear phase for large coupling, e.g. $\tilde{\alpha} > 20$. This regime is known as the strong back-reaction  
regime~\citep{Caravano:2022epk}. Furthermore, $\xi$ 
does not experience the oscillatory epoch discussed in ~\citep{Cheng:2015oqa,Domcke:2020zez,Caravano:2022epk, Peloso:2022ovc}, because before entering this phase at the end of previous stage, the next rolling field dictates the evolution of $\xi$. We work in the regime of negligible back-reaction such that the system never enter this phase and  the evolution of $\xi$ can not destroy the inflationary dynamics driven by the pseudo-scalar $\phi_a$.

A simple choice for the inflation potential  is the chaotic-type potentials ($V=\sum_{a=1}^{\sf N}m_a^2\Phi_a^2$)~\citep{Dimopoulos:2005ac,Yokoyama:2007dw,anber2006n} as in $\sf N$-flation {such that during the $a$-th stage only the field $\phi_a$ rolls down towards it potential minimum for some e-folds, oscillating rapidly at the bottom of its potential till its amplitude is effectively died out and the next field starts its rolling.}  However, the chaotic potentials is rule out by current Planck constraints~\citep{Planck:2018jri} even in the multi-field configuration~\citep{Wenren:2014cga,Easther:2005zr} due to the large tensor-to-scalar ratio value, $r_{\rm t}$. For example,  for the two-field and three-field ($\sf N=2, 3$) axion models with the pure chaotic potentials, our numerical results indicate that $r_{\rm t} \gtrsim 0.1$ which is in conflict with large scale CMB observations~\citep{Planck:2018jri}. One possible way to avoid this issue is to consider  the following simple potential form \citep{Kallosh:2018zsi,Kallosh:2022feu,Braglia:2020eai,Braglia:2020fms,Kallosh:2022vha}
\begin{equation}
	V(\Phi^a)=  V_{0}\dfrac{\Phi_1^2}{\Phi_1^2+m_1^2} +\sum_{a=\sf 2}^{\mathsf N} \dfrac{1}{2}m_a^2 \Phi_a^2 \,,
	\label{eq:potential}
\end{equation}
where $V_0$, $m_1$, and $m_a$ are constant parameters. In two-field case, this potential represents the well-known dilaton-axion inflation~\citep{Linde:2018hmx,Kallosh:2022feu}. We arrange that on CMB scales the first  pseudo-scalar field (the dilaton field) drive inflation while the remaining pseudo-scalar fields ($\sf N \ge 2$) with the standard chaotic type potential (now specifically called axionic fields) drive the rest of inflation.  Recently, in ~\citep{Braglia:2020eai,Braglia:2020fms,Kallosh:2022vha} the authors have shown that PBHs and GWs might be generated by considering a non-flat field space with the negative curvature in the absence of Chern-Simons coupling.  
In comparison, here we work with the flat field space while the instabilities induced  from the  Chern-Simons coupling are responsible for amplification of  power spectra.
Also note that instead of  potential  (\ref{eq:potential}) one may consider different examples as well. We only need to assume the first stage of inflation is driven with a potential different than simple chaotic potential  such that on CMB scales the value of $r_{\rm t}$ is small enough. 

After presenting the general setup, in the following  we consider two specific models:  dilaton-axion (model I) and dilaton-axion-axion (model II). Table~\ref{tab:models} presents the initial conditions and model parameters. The parameters are fixed to produce the correct COBE normalization at the CMB pivot scale $k_{\text{CMB}}=0.05 \hspace{1mm} \text{Mpc}^{-1}$~\citep{Planck:2018jri}. In addition, the parameter $\tilde{\alpha}_1$ for model I and $\tilde{\alpha}_2$ for model II have adopted the largest possible values consistent with the PBH bound~\citep{Garcia-Bellido:2016dkw} while $\tilde{\alpha}_1$ for model II has adopted the largest possible values consistent with the NANOGrav 11yrs data release \citep{NANOGRAV:2018hou}.

\begin{figure}
	\includegraphics[width=\linewidth]{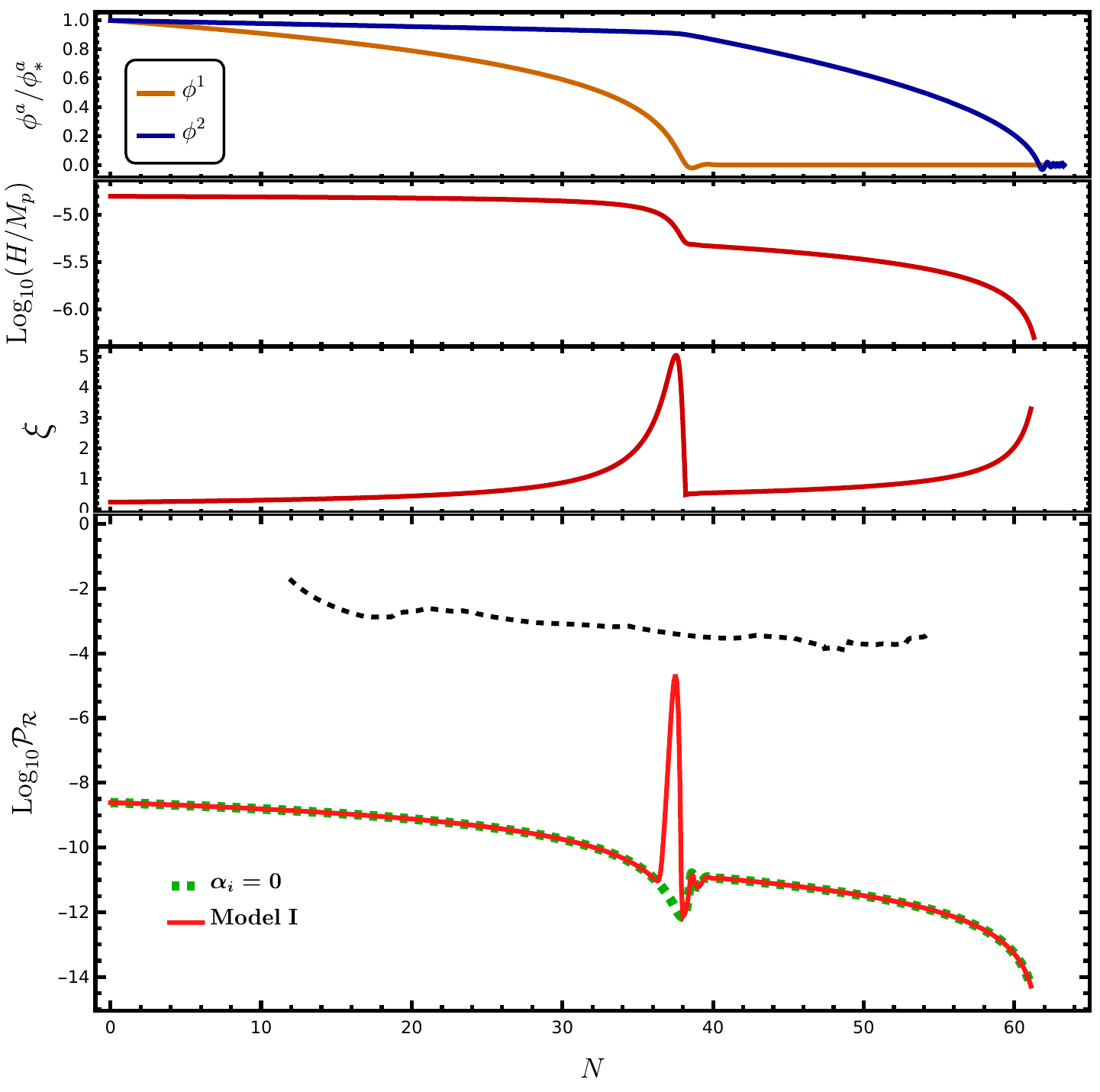}
	\caption{\scriptsize Evolution of the scalar fields, Hubble rate, instability parameter, and $\calP_\calR$ for the model I with two pseudo-scalars (one dilation and one axion). 
		The black dashed  curve presents the PBH bound~\citep{Garcia-Bellido:2016dkw}. A rise in $\xi$ yields to a localized peak 
		in $\calP_\calR$. 
		Note that the green dashed  curve shows that the power can not be enhanced  when $\tilde{\alpha}_i = 0$..
	}
	\label{fig:models1}
\end{figure}

\begin{figure}
	\includegraphics[width=\linewidth]{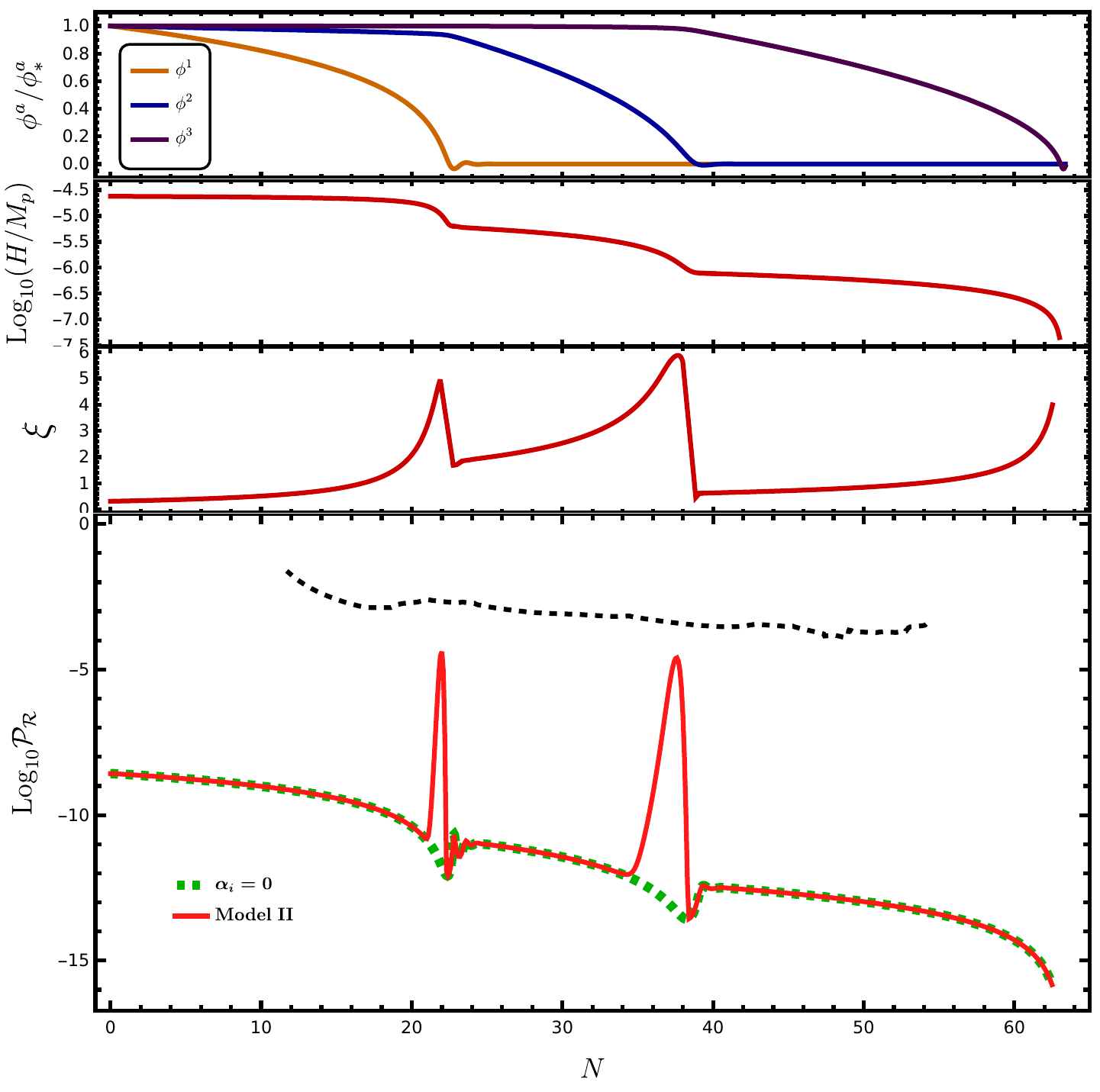}
	\caption{\scriptsize The same as in Fig. \ref{fig:models1} but for the model II. 
		%Evolution of the scalar fields, Hubble rate, instability parameter, and curvature power spectrum for the model II. 
		Since we have three pseudo-scalars (one dilaton and two axions) there are two peaks in evolution of $\xi$ yielding to two localized peaks in $\calP_\calR$.}
	\label{fig:models2}
\end{figure}

As shown in Figs.~\ref{fig:models1} and \ref{fig:models2}, the background experiences several inflationary phases in both models. The first inflationary phase is driven by dilaton $\Phi_1$  while the other fields remain frozen. After $\Phi_1$ has reached to its minimum and its energy is died out after a few rapid oscillations the axion fields drive the next inflationary phase each in turn. The evolution of the Hubble parameter is also presented in Figs.~\ref{fig:models1} and \ref{fig:models2} in accord with multiple inflationary phases. The numerical values of $\{N_{\rm end}, r_{\rm t}, n_{\rm s}\}$ for the total number of $e$-folding, the tensor-to-scalar ratio, and the spectral index on CMB scales   are $\{61.1, 0.014, 0.961\}$ for model I and $\{62.5, 0.032, 0.935\}$ for model II which are in close agreement with analytic results~\citep{Kallosh:2022feu}.

\begin{figure}
	\centering
	\includegraphics[width=\linewidth]{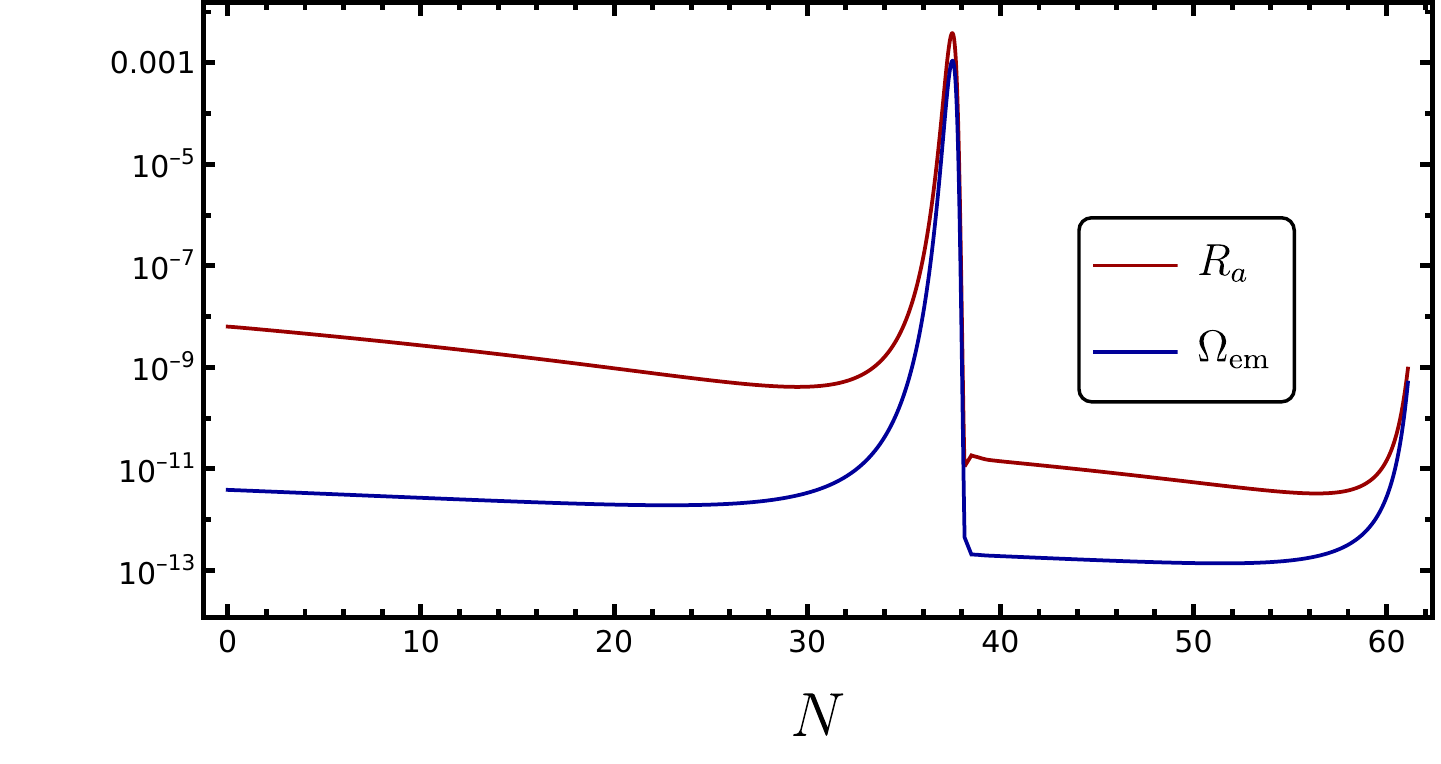}
\\
	\includegraphics[width=\linewidth]{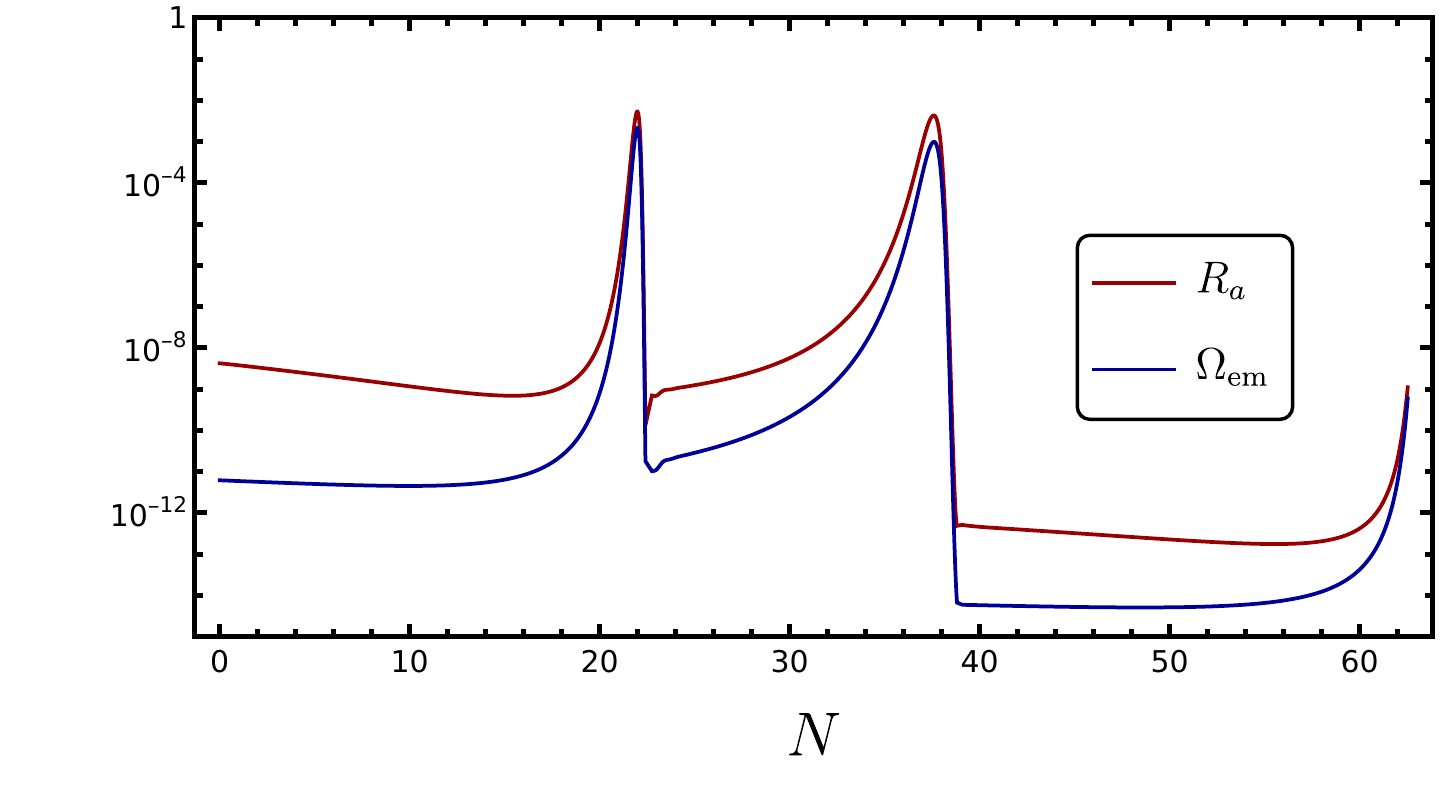}
	\caption{\scriptsize  Evolution of the back-reaction parameters for  model I (\textbf{top}) and model II  (\textbf{below}). We see that both of these parameters are small and the back-reactions effects are negligible.}
	\label{fig:backreaction}
\end{figure}

As mentioned earlier, we must ensure that the tachyonic growth of gauge field fluctuations does not modify the slow-roll inflationary dynamics. 
To do this, let us define the following back-reaction parameters 
\begin{equation}
R_{\mathrm{a}}\equiv |\frac{J_{\mathrm{a}}}{3 H \dot{\phi}_{0}}|\ll 1;  \hspace{0.5cm} \Omega_{\mathrm{ em}}\equiv \frac{\rho_{\mathrm{em}}}{3 \Mp^2 H^2} \ll 1,
\end{equation}
%\textbf{where} $\dot{\phi}_{0}^2=\delta_{ab} \dot{\phi}^{a} \dot{\phi}^{b}$ 
which measure the back-reaction effects on the dynamics of rolling fields and the total energy density, respectively. As shown in Fig.~\ref{fig:backreaction}, these dimensionless parameters are small so the evolution of $\xi$ does not destroy the inflationary dynamics driven by the pseudo-scalars for the parameter range which we work. 

\begin{deluxetable}{cccc}[th]
	\tablecolumns{3}
	\tablenum{1}
	%\tablewidth{0pt}
	\tabletypesize{\scriptsize}
	\tablehead{
		\colhead{Model} &
		\colhead{$(\tilde{\alpha}_1, \tilde{\phi}_*^1)$} &
		\colhead{$(\tilde{\alpha}_2, \tilde{\phi}_*^2,\tilde{m}_2)$} &
		\colhead{$(\tilde{\alpha}_3, \tilde{\phi}_*^3,\tilde{m}_3)$ }
	}
	\tablecaption{\scriptsize Model parameters $(\tilde{\alpha}_a,m_a)$ and initial conditions for pseudo-scalar fields, $\phi_*^a$ with $V_0 = 500 m_2^2 \Mp^2$ and $m_1=\sqrt{6} \Mp$ for both models and define the dimensionless parameters $\tilde{m}_a \equiv 10^{5}m_a/\Mp$ and $\tilde{\phi}_*^a \equiv \phi_*^a/\Mp$. The couplings to gauge fields $\tilde{\alpha}$ are chosen as large as allowed by the PHB bounds~\citep{Garcia-Bellido:2016dkw} and NANOGrav 11yrs data sets~\citep{NANOGRAV:2018hou} \label{tab:models}}
	\startdata
	I & (10.2,5.8) & (5,10.5,0.125) &  - \\
	II & (8.7,5) & (13.2,8,0.2) & (5,10,0.02)\\
	\enddata
\end{deluxetable}


\section{Curvature Perturbations and PBH Formation}
\label{sec:Curvature}
Now we study the  evolution of fluctuations in our setup. Because of the  coupling between the scalar fields and the gauge field, there will be source terms in the equation of motion of the scalar field fluctuations. The scalar field $\Phi^{b}$ can be decomposed into the background part $\phi^{b}$ and its canonical perturbation $\hat{Q}^{b}$ as $\Phi^{b}=\phi^{b}+{\bf a}\,  \hat{Q}^{b}$. In the spatially flat gauge, the equation of motion for the modes $\hat Q^{a}_k$ in momentum space reads \citep{Dimastrogiovanni:2018xnn,Linde:2012bt,Ozsoy:2017blg} 
\begin{eqnarray}\label{Qak}
	\bigg(\partial^2_\tau + k^2 -\frac{\textbf{a}''}{\textbf{a}}\bigg)\hat Q^a_k(\tau)+\textbf{a}^2 \nonumber \left(
	\dfrac{\partial^2 V}{\partial{\phi^{a}}\partial{\phi^{b}}}
	\right) \hat Q^b_k(\tau)=&&\\
	\frac{\tilde \alpha^{a} {\bf a}^{3}}{f} \int \frac{\dd ^3k}{(2\pi)^{3/2}}~e^{-i \bf{k} \cdot \bfx}~\vec{E}.\vec{B} \,,&&
\end{eqnarray}
where $\tau$ is the conformal time, $\dd \tau \equiv {\bf a}(t) \dd t$. The solution for $\hat Q^{a}$ can be separated into two uncorrelated parts $\hat Q_{a}=\hat{Q}_{a}^{(\text{v})}+\hat{Q}_{a}^{(\text{s})}$ where $\hat{Q}_{a}^{(\text{v})}$ represents the solution to the homogeneous part of Eq. \eqref{Qak} which reduces to Bunch-Davies vacuum on small scales, whereas $\hat{Q}_a^{(\text{s})}$ is the particular solution obtained by the Green function \citep{Barnaby:2011vw}. Also, since there is no interaction between the fields, the equations for $\hat Q^{a}$ are decoupled. 
Finally, the power spectrum of curvature perturbations, which is defined as $\mathcal{R}_k =\sum_{a} (H/{\bf a} \dot \phi^{a})\hat Q^{i}_k$ 
at horizon crossing time, $N_k$, becomes~\citep{Barnaby:2010vf,Barnaby:2011vw}
\begin{eqnarray}
	\label{powerspectra}
	{\cal P}_{\cal R}(k) \simeq \dfrac{H^2}{8\pi^2 \Mp^2 \epsilon_{_H}} %\left(\dfrac{k}{k_{\text{CMB}}}\right)^{n_s-1}
	\Big(
	1+\dfrac{H^2}{8\pi^2 \Mp^2 \epsilon_{_H}} f_{2}(\xi) e^{4\pi \xi}
	\Big) \,,
\end{eqnarray}
where $\epsilon_{_H}$ is the first slow-roll parameter and the dimensionless function $f_2(\xi)$ can be estimated for large $\xi$ as $10^{-5}/\xi^6$~\citep{Barnaby:2011vw}.
The first term in Eq. (\ref{powerspectra}) stands for the standard vacuum contribution to the power spectrum~\citep{Yokoyama:2007dw,Baumann:2009ds}.

The  curvature perturbations power spectrum for the models in Table~\ref{tab:models} are illustrated in Fig.~\ref{fig:models1} and \ref{fig:models2}. As can be seen, a rise in $\xi$ amplifies the  scalar power spectrum for the mode that leaves the Hubble radius at the transition time between the two stages. The location of $a$-th peak, $N_{a}$ (number of e-fold since the start of inflation), is given by the initial condition $\phi^a_*$, while the amplitude of ${\cal P}_{\cal R}(k)$ is controlled by $\tilde{\alpha}_a$.
Interestingly, for values of $\tilde{\alpha}_a$ considered in Table~\ref{tab:models}, the enhancement in the power spectrum is large enough to seed PBH formation due to the gravitational collapse of large density fluctuations after horizon re-entry during radiation-dominated era~\citep{Sasaki:2018dmp}. 

After PBH production, the next step is to determine the fraction of PBH abundance in dark matter density at the present epoch. It is roughly given by~\citep{Sasaki:2018dmp}
\begin{equation}
\label{fPBH}
	f_{\text{PBH}}(M_a) \simeq 2.7 \times 10^{8} \Big(%\frac{0.2}{\gamma}	\sqrt{\frac{g_{*}}{10.75}}
	\frac{M_{\odot}}{M_a}\Big)^{\frac{1}{2}} \beta(M_a) \,,
\end{equation} 
where the mass corresponding to $a$-th peak, $M_{a}$, can be estimated by the following relation as we assume an instant reheating at the end of inflation~\citep{Garcia-Bellido:2016dkw}.
\begin{align}
	\dfrac{M_{a}}{M_{\odot}} \simeq 10^{-13} \, %\big(\dfrac{\gamma}{0.2}\big) 
	\Big(\dfrac{10^{-6} \Mp \, H_{\rm end}}{{H_{a}^2}}\Big) \, 
	e^{2\left(N_{\rm end}-N_{a} - 22.25 \right)
		%(N_{\rm tot}-N_{\rm p})
	} \, ,
\end{align}
where $M_{\odot}$ is the solar mass and $H_{\rm end}$ and $H_a$ are the Hubble rates at $N_{\rm end}$ and $N_a$, respectively. Moreover, the $\beta$ is the mass fraction of PBHs at formation.

In the Press-Schechter formalism \citep{Press:1973iz}, $\beta$ is defined as the probability that the Gaussian comoving curvature perturbation $\mathcal{R}$ (or equivalently the density contrast $\delta$) is greater than a certain threshold value $\mathcal{R}_{c}$ (or $\delta_{c}$) for PBH formation ~\citep{Press:1973iz,Lyth:2012yp,Byrnes:2012yx,Garcia-Bellido:2016dkw}. Moreover, one can also compute $\beta$ by using the peak theory formalism where the primordial over density condition is expressed in terms of the peak value of a fluctuation mode \citep{young2014calculating,Musco:2018rwt,yoo2018primordial}, in contrast to the average value utilized in Press-Schechter theory. In addition,  in the peak theory formalism, the PBH formation probability is highly sensitive to the change in the tail of the fluctuation distribution. 

The threshold for PBHs at the cosmological horizon crossing 
has been widely computed in the literature  by making use of a linear extrapolation from the superhorizon regime. However, since the non-linear  relation between the density contrast and the curvature perturbation is neglected  it does not yield to the right amplitude of the perturbation at the cosmological horizon crossing \citep{Biagetti:2021eep,Musco:2018rwt}.

Taking into account this  non-linear effects, the mass fraction is defined as~\citep{Biagetti:2021eep}
\begin{align}
\label{beta-delta}
\beta \equiv \int_{\delta_{l,c}}^{4/3} \dd \delta_l ~ \kappa
\Bigg(
\delta_l-\dfrac{3}{8}\delta_l^2
-\delta_{c}
\Bigg)^{\tilde{\gamma}}~f_{\delta_l}(\delta_l)
\end{align}
where the probability distribution  $f_{\delta_l}(\delta_l)$ is given by \eqref{f_delta},  $\kappa=3.3$, and $\tilde{\gamma}=0.36$ for the collapse at the radiation-dominated epoch. Moreover, the field $\delta_{l}$ depends on the threshold density contrast through the following expression ( see App. \ref{PDF_delta} for more details).
\begin{align}
\delta_{l,c} = \dfrac{4}{3}\bigg(
1-\sqrt{1-\dfrac{3}{2}\delta_{c}}
\bigg)
\end{align}
in which one can take $\delta_{c} \simeq 0.59$ for a monochromatic curvature perturbation power spectrum~\citep{Musco:2018rwt, Musco:2020jjb}.

\begin{figure}
	\includegraphics[width=\linewidth]{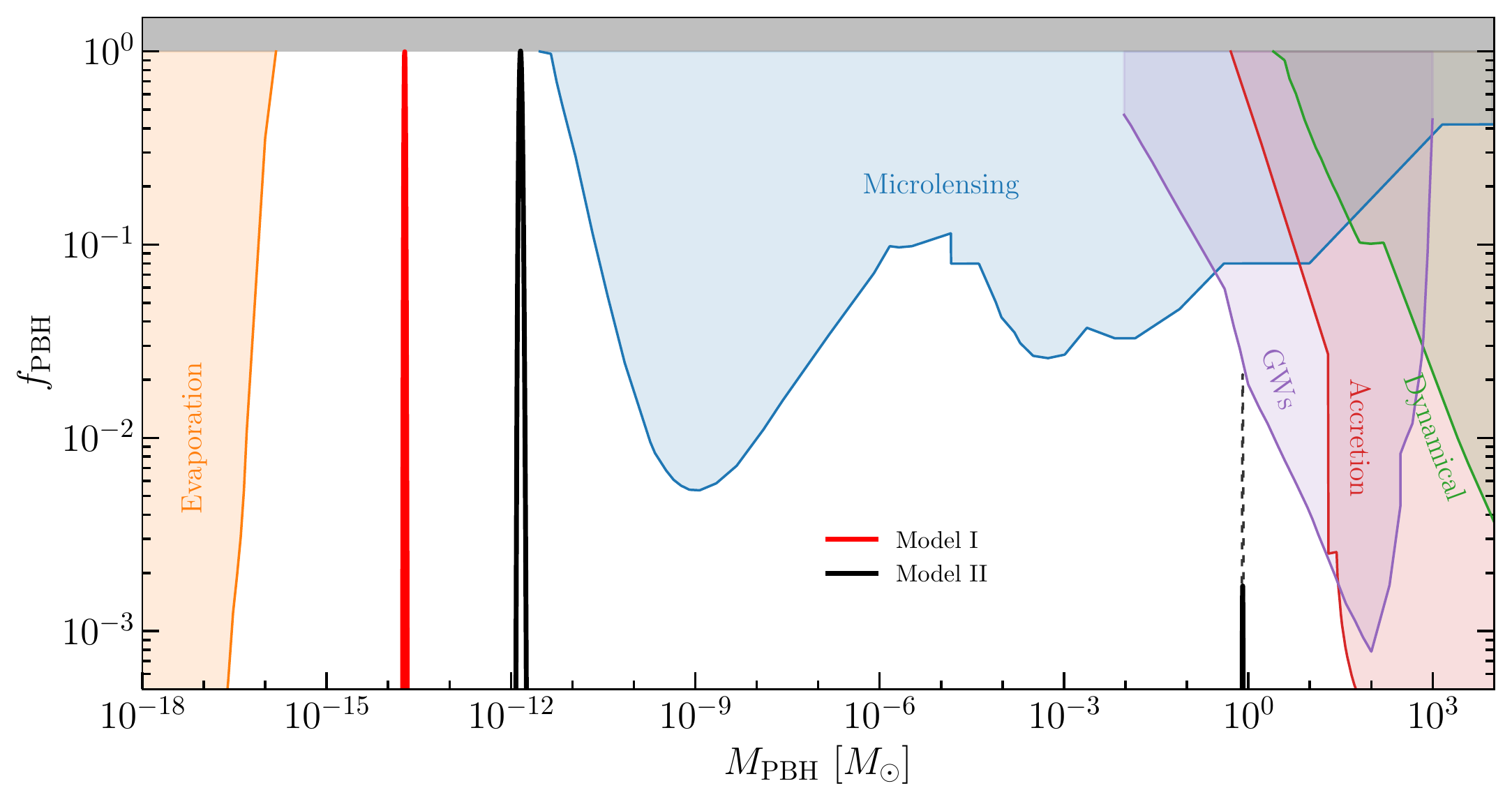}
	\caption{\scriptsize Fraction  $f_{\text{PBH}}$ as a function of the mass of the formed PBHs in unit of solar mass for models in Table.~\ref{tab:models}. The observational constraints are taken from Refs.~\citep{Green:2020jor,bradley_j_kavanagh_2019_3538999,Carr:2020gox}. The black dashed narrow band mass functions, corresponding to $M_{\rm PBH} \simeq 0.8M_\odot$ with $f_{\rm PBH} \simeq {\cal O}(0.1)$ for the case with $\tilde{\alpha}_1 = 8.9$. 
	}
	\label{fig:fPBH}
\end{figure}

In Fig.~\ref{fig:fPBH}, we have depicted $f_{\rm PBH}$ for the models introduced in Table~\ref{tab:models}.  As illustrated, the formed PBHs can furnish  a large fraction of total DM abundance. In particular, for model I and II, we obtain $f_{\rm PBH} \simeq 1$ corresponding to $M_{\rm PBH} \sim 10^{-14}M_\odot$ and $M_{\rm PBH} \sim 10^{-12}M_\odot$, respectively. Additionally, as illustrated in Fig.~\ref{fig:models2}, because the first peak in the scalar power spectrum for model II is not amplified large enough, its corresponding PBH mass, i.e. $M_{\rm PBH} \sim M_\odot$ can not contribute significantly to the dark matter density in the universe today. However, in the case with $\tilde{\alpha}_1 = 8.9$ and the other initial conditions like model II, one obtains $f_{\rm PBH} \simeq {\cal O}(0.1)$ for almost the same PBH mass. In spite of this fact, such a model conflicts with the recent observational constraint on the GWs determined by the NANOGrav 11yrs data release~\citep{NANOGRAV:2018hou}.

%
%In Appendix \ref{PDF_delta}, we also attempt to compare above results with those coming from the PBH mass function $\beta$ computed by using the PDF for the density contrast which is non-linearly related to the comoving curvature perturbation~\citep{Sasaki:2018dmp,Musco:2018rwt,Hooshangi:2021ubn,Hooshangi:2022lao}. 

\section{Primordial and Induced GWs}
\label{sec:GW}
In addition to the quantum vacuum fluctuations of metric during inflation, there are two distinct populations of stochastic GWs  in our inflationary scenario. The first contribution is related to the GWs generated form the amplified
gauge fields during inflation~\citep{Barnaby:2010vf,Sorbo:2011rz,Namba:2015gja,Ozsoy:2020ccy,Ozsoy:2020kat}. The second
contribution is the so-called \textit{induced}
GWs  originating from the enhanced second order scalar fluctuations~\citep{10.1143/PTP.37.831,Matarrese:1992rp,Matarrese:1993zf,Matarrese:1997ay,Ananda:2006af,Baumann:2007zm,Espinosa:2018eve,Kohri:2018awv,Domenech:2021ztg}.

%As mentioned, the tachyonic gauge fields can source large tensor fluctuations. 
Due to the parity-violating nature of the system, the right and left helicities of tensor modes have different amplitudes~\citep{Sorbo:2011rz}. The equation of motion for two  canonical tensor helicity $\hat{h}^{\lambda}$ is given by
\begin{align}
	\label{tensormode}
	\big(\partial^2_\tau + k^2 -\dfrac{2}{\tau^2}\big)\hat h^{\lambda}_k(\tau) &=
	\frac{-\textbf{a}^{3}}{\Mp} \Pi_{ij}^{\lambda}(\textbf{k}) \times \nonumber\\
	\int \frac{\dd^{3}k \,\ e^{- i  \textbf{k}. \textbf{x}}}{(2 \pi)^{3/2}}&  ~\big[E_{i}E_{j}+B_{i}B_{j}\big] \,,
\end{align}
where $\Pi^{\lambda}_{ij}$ is the transverse traceless projector~\citep{Sorbo:2011rz}.
Similar to scalar fluctuations, we decompose $\hat h^{\lambda}$ into a vacuum mode, $\hat h_{\lambda}^{(\text{v})}$, and the sourced mode, $\hat h_{\lambda}^{(\text{s})}$. Adding up these two contributions, one finds the tensor power spectrum for each polarizations mode to be \citep{Sorbo:2011rz}    %~\citep{Barnaby:2011vw} 
\begin{equation}
	\label{eq:primordialGW}
	\mathcal{P}_{\lambda}^{(\rm p)}(k)\simeq\dfrac{H^2}{\pi^2 \Mp^2} 	\big(1+\dfrac{2H^2}{\Mp^2} f_{ \lambda}(\xi) e^{4\pi \xi}\big) \,,
\end{equation}
where the superscript ${(\rm p)}$ stands for the primordial contribution in which the first term represents the contribution from vacuum fluctuations. Moreover, the dimensionless function $f_{ \lambda}(\xi)$ at large $\xi$ for the right and left helicities are approximately $10^{-7}/\xi^{6}$ and $10^{-9}/\xi^{6}$, respectively~\citep{Sorbo:2011rz}.  Correspondingly, the main contribution to the primordial tensor power spectrum, $\mathcal{P}^{(\rm p)}_{h}=\sum_{\lambda} \mathcal{P}_{\lambda}^{(\rm p)}$, comes from the right helicity GW modes ~\citep{Sorbo:2011rz,Barnaby:2011qe}.

As stated earlier, the  GWs can be induced from the amplified curvature perturbations in Eq. \eqref{powerspectra}~\citep{Zhou:2020kkf,Domenech:2020kqm,Pi:2020otn,Cai:2019amo,Ozsoy:2020kat}. Indeed, the large second order scalar fluctuations on small scales induce tensor perturbations after the horizon re-entry during radiation-dominated era. With regard to the population of GWs discussed above, one deals with  multiple integrals of the following form ~\citep{Kohri:2018awv}
\begin{eqnarray}
	{\cal P}_{h}^{\rm (ind)}\!\sim \!\int \!\dd k \!\int \!\dd k' \bigg[\!\int \!f(k,k',t)\dd t
	\bigg]^2 {\cal P}_{\cal R}(k){\cal P}_{\cal R}(k')~~~~
	\label{eq:induGWs}
\end{eqnarray}
where $f(k,k',t)$ is an oscillating function and $t$ describes the time when the GW is sourced from the scalar modes~\citep{Kohri:2018awv,Cai:2018dig,Acquaviva:2002ud,Inomata:2016rbd,Domenech:2021ztg}.  
Finally, the total present-day energy density of GWs   is given by~\citep{Ozsoy:2020kat,Baumann:2007zm,Espinosa:2018eve}
\begin{align}
	\Omega_{\rm GW}(k) =  \Omega^{(\rm p)}_{\rm GW}(k) +\Omega^{(\rm ind)}_{\rm GW}(k) \,,
\end{align}
in which $\Omega^{(\rm p)}_{\rm GW}(k) $ and $\Omega^{(\rm ind)}_{\rm GW}(k)$
represent the fraction energy density of primordial GWs induced by the tachyonic gauge field mode and  the induced GWs from the second order scalar 
perturbations respectively.

\begin{figure}		\includegraphics[width=\linewidth]{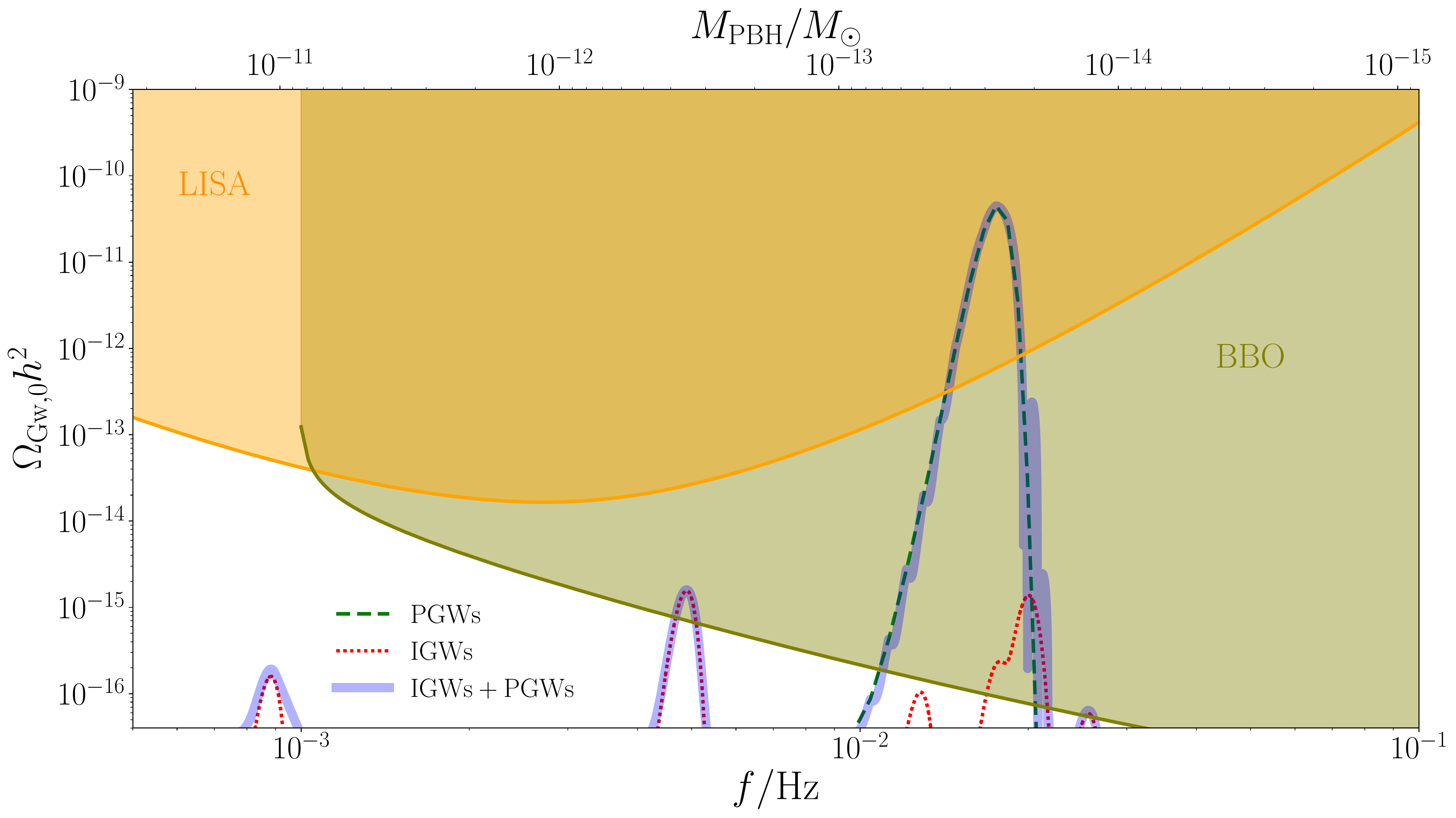}
	\caption{\scriptsize The energy density of the GWs for model I with respect to frequency. The contributions of primordial (polarized) and induced GWs are plotted separately as well. 
	The shadowed regions represent the sensitivity curves of various GW detectors~\citep{Schmitz:2020syl,schmitz_kai_2020_3689582}.  The largest peak of the curve is due to primordial polarized  GWS given in  \eqref{eq:primordialGW} while the oscillations in the curve originate from convolution integrals in  Eq. \eqref{eq:induGWs}. }
	\label{GWplot1}
\end{figure}

\begin{figure}		\includegraphics[width=\linewidth]{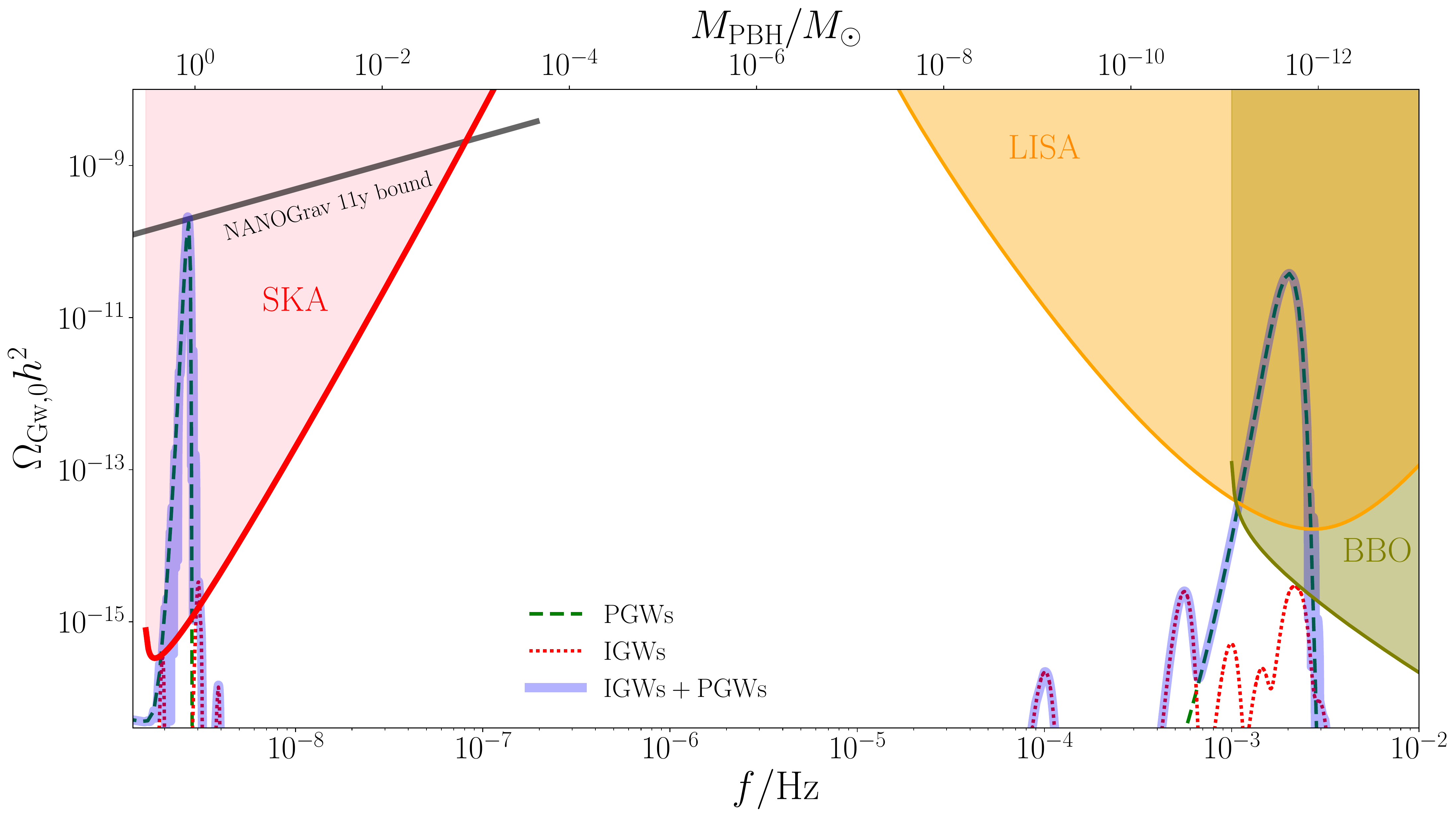}
	\caption{\scriptsize The same as Fig.~\ref{GWplot1} but for model II. Note that the model II with two rolling axions yields to two separated sets of curves. The black solid line  represent the observational limits imposed by the NANOGrav 11yrs data release~\citep{NANOGRAV:2018hou}. }
	\label{GWplot2}
\end{figure}

In Figs. \ref{GWplot1} and \ref{GWplot2}, we have plotted the quantity $\Omega_{\rm GW} h^2$ (light blue curve) which is the sum of $\Omega_{\rm GW}^{(\text{p})} h^2$ (green dashed curve), and $\Omega_{\rm GW}^{(\text{ind})} h^2$ (red dotted  curve) against the frequency with $h^2=0.49$ together with the sensitivity of the various forthcoming GW experiments \textit{e.g.} the %Laser Interferometer Space Antenna (LISA) 
LISA \citep{Bartolo:2016ami}, %the Big Bang Observatory (BBO)
BBO \citep{Crowder:2005nr,Corbin:2005ny,Baker:2019pnp}, 
%the Square Kilometre Array (SKA)
SKA~\citep{Carilli:2004nx,Janssen:2014dka,Weltman:2018zrl}, and 
%the Parkes Pulsar Timing Arrays (PPTA)
PPTA~\citep{Manchester:2012za,Shannon:2015ect}.
As can be seen, the summit of the total GW curves are related to $\Omega_{\rm GW}^{(\text{p})} h^2$, while $\Omega_{\rm GW}^{(\text{ind})} h^2$ constitutes a sub-dominant portion of the total GW signal around the biggest peak. In addition, 
the oscillations in the curves originate from convolution integrals in  Eq. \eqref{eq:induGWs}.

Clearly, for the models I, $\Omega_{\rm GW} h^2$ falls within the sensitivity of the BBO and peaks well inside the range of detectability of LISA. Remarkably, for the model II, we observe that the double rises in GWs are
detectable by LISA and SKA. A similar feature has been observed in~\citep{Bhaumik:2022pil} as a signal of a non-thermal baryogenesis from evaporating PBHs. 

On the other hand, the current severe constraint on stochastic GW background in nHZ regime, i.e. NANOGrav 11yrs~\citep{NANOGRAV:2018hou} can put restrictions on the parameters of our model. We observe from Fig. \ref{GWplot2} that the first peak of $\Omega_{\rm GW} h^2$ for model II with $\tilde{\alpha}_1 = 8.7$ is located at SKA scales by respect to the NANOGrav 11yrs bound. However, for the model with $\tilde{\alpha}_1 = 8.9$, the result clashes with the NANOGrav 11yrs constraint. Generally, the both observational data of PBH limits and GWs should be considered together to limit the parameter space of our model. 

As a final remark, the sign of $\tilde{\alpha}_{a}$
determines which polarization of the gauge field is amplified and hence we have considered a simple setup where only one polarization is amplified all the time. Nevertheless, when we switch the signs of $\tilde{\alpha}_{a}$, then the other polarization gets amplified as well and we must perform the mode analysis more carefully. Moreover, since the peaks in the GW spectrum are so separated, one can  observe only one at a time, so the sign of the chirality of the other, non observable ones ( at frequencies that are too high/too low)  will not matter. 
We leave a comprehensive analysis on this general situation for a future work.


\vspace{1cm}
\section{Summary and Discussions}
\label{sec:Summary}
We have studied a model of inflation with multiple pseudo-scalar fields coupled 
to a gauge field via the Chern-Simons type interactions.   There are multiple stages of inflation driven by each scalar field. To evade the constraint on tensor-to-scalar ratio, we have considered a setup where the first stage is driven by a dilaton field while the remaining stages of inflation below CMB scales are driven with multiple axionic fields with the standard chaotic type potentials.  However, our setup can be extended to more complicated potentials, such as  $\alpha$-attractor model \citep{Kallosh:2022feu}. The enhanced power spectrum from the gauge field instability 
can generate PBHs with various masses  which can furnish a large fraction of total DM while satisfying the bounds on PBHs formation \citep{Garcia-Bellido:2016dkw}.  In addition, GWs can be generated both from second order scalar perturbations as well as from the tachyonic gauge field perturbations  with  distinct features on the location of the peaks and their oscillatory behaviours. These signals  are within the detection range of the future GW  observatories. %which may put constraint on model parameters. 
There are a number of directions in which the current investigations can be extended. These include investigating the non-Gaussianity of the perturbations \citep{DAmico:2021fhz} and its effects on the induced GWs~\citep{Adshead:2021hnm,DAmico:2021vka} and the PBHs formation~\citep{Biagetti:2021eep}.

\begin{acknowledgments}
This work is dedicated to the memory of Prof. Mohammad Reza Setare (1974-2022). We are grateful to Lorenzo Sorbo for useful comments on the draft. We acknowledge the partial support from the ``Saramadan" federation of Iran. A. T. would like to thank Yukawa Institute for Theoretical Physics (YITP) for their kind hospitality during the development of this work. We would like to thank the 
anonymous referee for the insightful comments and suggestions  which have improved the quality of the paper.
\end{acknowledgments}

\appendix
\section{\textbf{PDF of the density contrast}}
\label{PDF_delta}

The main aim of this appendix is to derive the probability distribution function (PDF) for the smoothed density contrast and hence compute the mass fraction $\beta$. By following the same methodology proposed in~\citep{Biagetti:2021eep}, we first consider the smoothed density contrast, $\delta_m$, as ~\citep{Biagetti:2021eep,Musco:2018rwt,young2019primordial}
\begin{align}
\label{delta_m}
\delta_m = \delta_l- \dfrac{3}{8}\delta_l^2 \,;
\hspace{1cm}
\delta_l \equiv - \dfrac{4}{3} r_m {\cal R}'(r_m)\,, 
\end{align}
where the prime here denotes the derivative with respect to the radial coordinate $r_m$ which indicates the location of the maximum of the compaction function. More importantly,  the relation \eqref{delta_m} shows non-linear relation between the smooth density contrast and spatial derivative of the curvature perturbation\footnote{Due to the dependence of the field $\delta_{l}$ on the derivative of the comoving curvature perturbation, one can always add or subtract to the comoving curvature perturbation a constant by a coordinate transformation on superhorizon scales and this may not affect on any physical result.}. Then, 
thanks to the conservation of the probability, we compute the PDF of the auxiliary field $\delta_l$, indicates by $f_{\delta_l}$, in order to calculate the mass fraction \eqref{beta-delta}. 

It is worthwhile noting that, in our case of study, sourced scalar fluctuations $\hat{Q}_a^{(\rm s)}$ originate from the convolution of two Gaussian gauge fields and hence the PDF of curvature perturbation obeys a $\chi^{2}$ statistics \citep{Linde:2012bt}. Therefore, we can consider~\citep{Linde:2012bt}
\begin{align}
{\cal R} = g^2 - \langle g^2 \rangle
\end{align}
where $g$ stands for the Gaussian distributed field\footnote{ The stochastic properties of the gauge field $A$ are close to those in a free theory, namely it has Gaussian perturbations
around $\langle A \rangle = 0$.} whose PDF is given by
\begin{align}
f_g(x) = \dfrac{1}{\sqrt{2\pi}\sigma_g} \exp(-\frac{x^2}{2\sigma_g^2})
\end{align}
in which $\sigma_g^2 \equiv \langle g^2 \rangle$ is the variance. Having the above PDF allows us to find $\langle g^4 \rangle = 3\sigma_g^4$ and hence $\sigma_{\cal R}^2 \equiv \langle {\cal R}^2 \rangle = 2\sigma_g^4$. For simplicity we can consider the power spectra of curvature perturbations shown in Fig.~\ref{fig:models1} 
%and \ref{fig:models2} 
as a monochromatic spectra ${\cal P}_{\cal R}(k) \simeq {\cal P}_* k_* \delta_D(k-k_*)$ (where $\delta_D$ is the Dirac Delta function) peaked at a momentum scale $k_*$ such that $r_m k_* \simeq 2.74$~\citep{Musco:2018rwt}. Therefore, we have
\begin{align}
\sigma_{\cal R}^2 \equiv \int ~ \dd \ln k~{\cal P}_{\cal R}(k) \simeq {\calP}_* \,.
\end{align}
By taking into account the above result, one can relate the variance of Gaussian field $g$ to the amplitude of peak of the curvature perturbation, namely
\begin{align}
\sigma_g \simeq \big(
{\cal P}_*/2
\big)^{1/4} \,.
\end{align}
In addition, because $\langle g^2 \rangle$ is independent of $r_m$, one  obtains
\begin{align}
\delta_l = - \dfrac{8}{3} r_m g g'\,. 
\end{align}
Let us now consider the two uncorrelated fields $g$ and $g'$ which are both Gaussian random fields with the zero mean value. The variance of $g'$ is denoted as $\sigma_{g'} = k_* \sigma_g$. According to the conservation of the probability, the PDF of the density contrast is simply given by
\begin{align}
\label{f_delta}
f_{\delta_l} (\delta_l) &=\int ~ \dd g ~\dd g'~ \delta_D\Big(\delta_l+\dfrac{8}{3} r_m g g'\Big) f_g(g)~ f_{g'}(g') \,  \nonumber 
\\
&= \dfrac{3}{8r_m}\int ~ \dd g ~ \dfrac{f_g(g)}{\abs{g}}~ f_{g'}\Big(\dfrac{-3\delta_l}{8gr_m}\Big)  \,  \nonumber 
\\
&= \alpha~ K_0\big(\alpha~\abs{\delta_l}\big) \,,
\end{align}
where $K_n(x)$ is the modified Bessel function of the second kind and 
\begin{align}
\alpha \equiv \dfrac{3}{8\pi r_m \sigma_g \sigma_{g'}} = \dfrac{3\sqrt{2}}{8\pi k_*r_m \sqrt{{\cal P}_*}} 
\end{align}

Through the use of PDF \eqref{f_delta}, we are able to compute the PBH mass fraction in Eq. \eqref{beta-delta} and then estimate the fractional abundance of PBHs at the present epoch via Eq. \eqref{fPBH}. 

 Finally, it is worth emphasizing  the importance of the nonlinear effects of curvature perturbation, as given in Eq.  \eqref{delta_m}, for calculating $f_{\rm PBH}$ in our setup. To do so, we consider the conventional method for estimation the mass fraction $\beta$. The method is based on the relation
\begin{align}
\beta \sim \gamma \int_{{\cal R}_c}^{\infty}~\dd {\cal R}~ f_{\cal R}({\cal R})
\end{align}
where $\mathcal{R}_{c}\simeq 1$~\footnote
%%%%%%%%%%%%%%%%%%%%%%%%%%%
{Recent numerical and theoretical investigations imply that $\mathcal{R}_{c} \sim \mathcal{O}(1)$ ~\citep{musco2005computations,musco2009primordial,nakama2014identifying,harada2013threshold}.  Moreover, the proper value of the threshold depends on the shape of the power spectrum of the curvature perturbation. In this work, we consider $\mathcal{R}_{c} \sim 1.75$ according to the value of density threshold $\delta_{c} \sim 0.55$ quoted in \citep{musco2021threshold} by making  use of the linear relation $\mathcal{R}_{c}=9/(2 \sqrt{2})\delta_{c}$ between curvature and density threshold \citep{drees2011running,young2014calculating,motohashi2017primordial}.} 
%%%%%%%%%%%%
is the threshold value and $f_{\cal R}$ is the PDF of the curvature perturbations and the value of the constant of proportionality $\gamma \simeq 0.2$ is proposed in ~\citep{Carr:1975qj,Sasaki:2018dmp}. In our case of study, sourced scalar fluctuations $\hat{Q}_a^{(\rm s)}$ originate from the convolution of two Gaussian gauge fields and hence the PDF of curvature perturbation obeys a $\chi^{2}$ statistics \citep{Linde:2012bt}.
Consequently, the fraction $\beta$ is related to power spectrum of curvature perturbation by ~\citep{Garcia-Bellido:2016dkw}
\begin{align}
\label{beta_R}
\beta(N_{a}) \simeq {\rm Erfc}\Bigg(
\sqrt{\dfrac{1}{2}+\dfrac{{\cal R}_c}{\sqrt{2{\cal P}_{\cal R}(N_a)}}}
\Bigg)
\end{align}
in which ${\rm Erfc} (x) \equiv 1-{\rm Erf} (x)$ is the complementary error function. Using \eqref{beta_R} for for the mass fraction in \eqref{fPBH}, the maximum value obtained for Table.~\ref{tab:models} corresponds to model I with $f_{\rm PBH} \sim {\cal O}(10^{-91})$! This estimate shows that the nonlinear effects of the curvature perturbation play a significant role in generating PBH in this class of models.

%%%%%%%%%%%%%%%%%%%%%%%%%%%%%%%%%%%%%%%%%%%%%%%

\bibliography{Multi_axion}{}
\bibliographystyle{aasjournal}

\end{document}